\documentclass{aa}
\usepackage{graphicx}
\usepackage[varg]{txfonts}
\usepackage{cases}
\usepackage{natbib}
\usepackage{color}

\begin{document}

\title{Quasi-periodic pulsations with periods that change depending on whether the
pulsations have thermal or nonthermal components}

\author{D.~Li \inst{1,2}, Q.~M.~Zhang \inst{1}, Y.~Huang\inst{1}, Z.~J.~Ning\inst{1}, and Y.~N.~Su\inst{1}}

\institute{Key Laboratory for Dark Matter and Space Science, Purple Mountain Observatory, CAS, Nanjing 210008, PR China \email{lidong@pmo.ac.cn} \\
           \and Key Laboratory of Modern Astronomy and Astrophysics (Nanjing University), Ministry of Education, Nanjing 210093, PR China}

\date{Received; accepted}

\titlerunning{QPPss with periods that change depending on whether the
pulsations have thermal or nonthermal components}
\authorrunning{D. Li et al.}

\abstract {Quasi-periodic pulsations (QPPs) typically display
periodic and regular peaks in the light curves during the flare
emissions. Sometimes, QPPs show multiple periods at the same
wavelength. However, changing periods in various channels are rare.}
{We report QPPs in a solar flare on 2014 October 27. They showed a
period change that depended on whether thermal or nonthermal
components were included. The flare was simultaneously observed by
many instruments.} {Using the fast Fourier transform (FFT), we
decomposed the light curves at multiple wavelengths into slowly
varying and rapidly varying signals. Then we identified the QPPs as
the regular and periodic peaks from the rapidly varying signals. The
periods are derived with the wavelet method and confirmed based on
the FFT spectra of the rapidly varying signals.} {We find a period
of $\sim$50~s from the thermal emissions during the impulsive phase
of the flare, that is, in the soft X-ray bands. At the same time, a
period of about 100~s is detected from the nonthermal emissions,
such as hard X-ray and microwave channels. The period ratio is
exactly 2.0, which might be due to the modulations of the magnetic
reconnection rate by the fundamental and harmonic modes of
magnetohydrodynamic waves. Our results further show that the
$\sim$100~s period is present over a broad wavelength, such as hard
X-rays, extreme-UV/UV, and microwave emissions, indicating the
periodic magnetic reconnection in this flare.} {To our knowledge,
this is the first report about period changes from thermal to
nonthermal components in a single flare that occur at almost the
same time. This new observational finding could be a challenge to
the theory of flare QPPs.}


\maketitle

\section{Introduction}
Quasi-periodic pulsations (QPPs) are very common features in
flare emissions. They typically exhibit periodic peaks in the light
curves. All these peaks have similar lifetimes, which results in
regular intervals between the peaks. The observed periods can
last from
milliseconds through seconds to minutes
\citep[e.g.,][]{Hoyng76,Karlicky05,Tan10,Su12,Ning14a,Li15,Zhang16}.
Previous observations indicated that the periods of these flares are strongly
dependent on the major radius of flare loops \citep{Aschwanden98}.
On the other hand, QPPs can be detected in intensity (or
flux) observations at wide bands, that is, in X-rays
\citep{Hoyng76,Dolla12,Ning14a,Hayes16} from extreme-ultraviolet
(EUV) \citep{Su12} to microwave \citep{Mangeney89,Tan10} emissions. QPPs
can also be observed  in Doppler shifts in spectral
observations, such as the hot ($>$6~MK) lines observed by {\it
SOHO}/SUMER \citep{Ofman02,Wang02}, the coronal lines from {\it
Hinode}/EIS \citep{Tian11}, and the flare and transition lines from {\it
IRIS} observations \citep{Li15,Liting15,Tian16}. Sometimes, QPPs
in the same flare display multiple periods \citep{Inglis09}, and these
multi-periods are often detected at the same channels, that
is, in sof X-ray (SXR)
\citep{Chowdhury15}, EUV \citep{Su12,Tian16}, H$\alpha$
\citep{Srivastava08,Yang16}, and microwave
\citep{Melnikov05,Kupriyanova14} channels. Multi-periods are
also found at various channels but in different flare loops
\citep{Zimovets10} or phases \citep{Hayes16,Tian16}. All these
period ratios are always deviating from 2.0.

Today, the formation mechanism of QPPs is still debated
\citep[e.g.,][]{Aschwanden87,Nakariakov09,Van16}. Generally, QPPs
are thought to be related with waves or energetic particles
(electrons). Magnetohydrodynamic (MHD) waves are most frequently
used to explain
QPPs. These include slow magnetoacoustic, fast kink, and sausage
waves \citep{Nakariakov09,Tan10,Su12}. QPPs can also be explained by
quasi-periodic magnetic reconnection
\citep{Nakariakov09,Li15,Liting15}. The periods can be spontaneous
\citep{Kliem00,Murray09} or may be modulated by some MHD waves
\citep{Chen06,Nakariakov11,Li15}.

Until now, QPPs in a flare that have periods that change in
different channels have rarely been simultaneously observed with
different instruments. We here analyze QPPs with changing periods in
a solar flare on 2014 October 27, which was detected by {\it The
Reuven Ramaty High Energy Solar Spectroscopic Imager} ({\it RHESSI})
\citep{Lin02}, the Atmospheric Imaging Assembly (AIA)
\citep{Lemen12} onboard {\it Solar Dynamics Observatory} ({\it
SDO}), {\it Nobeyama Radio Polarimeters} ({\it NoRP})
\citep{Nakajima85}, and the {\it Nobeyama Radioheliograph} ({\it
NoRH}) \citep{Hanaoka94} in the SXR, HXR, EUV/UV, and microwave
channels.

\section{Observations and results}

\begin{figure}
\centering
\includegraphics[width=\linewidth,clip=]{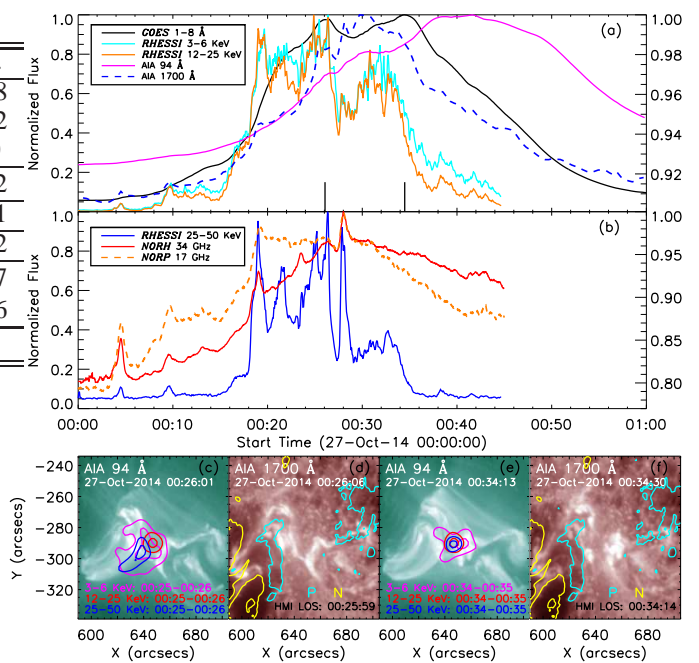}
\caption{Upper Panel: The normalized light curves of the flare on
2014 October 27 from {\it GOES} 1$-$8~{\AA}, {\it RHESSI} 3$-$6 keV
and 12$-$25 keV, {\it SDO}/AIA 94~{\AA} and 1700~{\AA}. The vertical
ticks indicate the time of two {\it GOES} SXR peaks. Middle panel:
The normalized light curves of the same flare from {\it RHESSI}
25$-$50 keV, {\it NoRP} 17~GHz, and {\it NoRH} 34~GHz. Lower panels:
{\it SDO}/AIA images at 94 (c, e) and 1700~{\AA} (d, f) at two SXR
peak times. The X-ray contour levels from the {\it RHESSI} images are
95\%, 80\%, while the HMI contour levels are at $\pm$800~G.
\label{light}}
\end{figure}

We explore an M7.1 flare that erupted in NOAA AR12192 (S12W42). It
started at $\sim$00:06~UT, peaked at $\sim$00:34~UT, and stopped at
$\sim$00:44~UT according to the {\it GOES} SXR flux. This flare was
detected by many instruments over a broad channel with various
temporal resolution, as shown in Table~\ref{tab}. Figure~\ref{light}
displays the normalized light curves and images detected by {\it
GOES}, {\it RHESSI}, {\it SDO}, {\it NoRP}, and {\it NoRH}. The
black profile in panel~(a) shows the {\it GOES} SXR flux at
1$-$8~{\AA}, which clearly shows two peaks at about 00:26~UT and
00:34~UT, indicating two heating processes during the impulsive
phase of the flare \citep[see also][]{Chen15,Polito16}. Panel~(a)
also displays the light curves from {\it RHESSI} 3$-$6~keV
(turquoise) and 12$-$25~keV (orange), AIA 94~{\AA} (purple), and
1700~{\AA} (blue). Panel~(b) shows the HXR flux at 25$-$50 keV
(blue), microwave emissions from {\it NoRP} 17~GHz (orange), and
{\it NoRH} 34~GHz (red). These three light curves in panel~(b) seem
to be well correlated and have similar peaks during the impulsive
phase of the flare; see also Table~\ref{tab}. The reference channel
is HXR emissions at 25$-$50~keV. Table~\ref{tab} gives the
correlation coefficients (c.c.) between the light curves at HXR
25$-$50~keV and other channels. Most of the coefficients are higher
than 0.8. However, the coefficient between the flux at AIA~94~{\AA}
is slightly lower, only about 0.57. The reason might be that the
heating process at AIA~94~{\AA} occurs later than that at other
channels. Figure~\ref{light}~(a) also shows that the peak time of
AIA 94~{\AA} is later than others, that is, later than {\it GOES}
1$-$8~{\AA} and AIA 1700~{\AA}. In order to clearly show the light
curves, the solid profiles in panels~(a) and (b) work with the left
axes, while the dashed profiles correspond to the right axes.

\begin{table}
\caption{Details of the observational instruments and data.}
\centering
\begin{tabular}{c c c c c c c}
 \hline\hline
Instruments    &  Channels           &   Cadence (s)     &  Descriptions  &  c.c.     \\
 \hline
               & 3$-$6 keV        &    4.0            & SXR            &  0.88     \\
{\it RHESSI}   & 12$-$25 keV      &    4.0            & SXR/HXR        &  0.92     \\
               & 25$-$50 keV      &    4.0            & HXR            &  1.0       \\
\hline
{\it GOES}     &  1$-$8 {\AA}     &   $\sim$2.0       & SXR            &  0.82     \\
\hline
{\it NoRP}     &  17 GHz          &    1.0            & Microwave      &  0.81     \\
\hline
{\it NoRH}     &  34 GHz          &    1.0            & Microwave      &  0.82      \\
\hline
               &   94 {\AA}       &    12             & EUV            &  0.57     \\
{\it SDO}/AIA  &   1700 {\AA}     &    24             & UV             &  0.86     \\
  \hline
{\it SDO}/HMI  &   6173 {\AA}     &    45             & magnetogram    &   --       \\
 \hline\hline
\end{tabular}
\label{tab}
\end{table}

Figure~\ref{light}~(c)$-$(f) shows the AIA images with a size of about
115\arcsec$\times$105\arcsec\ at 94~{\AA} and 1700~{\AA} from two
peak times marked by the vertical ticks in panel~(a). The light
curves at 94 and 1700~{\AA} in panel~(a) are integrated from the
all flare regions. The color contours in panels~(c) and (e)
represent the {\it RHESSI} SXR/HXR images using the CLEAN algorithm
from detectors 3, 4, 5, 6, and 8, while the color contours in
panels~(d) and (f) are from the {\it SDO}/HMI line-of-sight (LOS)
magnetograms at a scale of $\pm$800~G for positive (turquoise) and
negative (yellow) fields, respectively. The imaging observations
show that the flare has complex ribbons and the HXR sources are
located between two of them. This M7.1 flare is also associated with
strong magnetic fields that were connected by the positive and negative
poles.

Figure~\ref{light}~(b) shows several peaks during the impulsive
phase of the flare in HXR and microwave light curves between
00:18~UT and 00:28~UT. These peaks appear to be regular and
periodic, but they are superposed by the gradual background
emissions, which makes them difficult to distinguished. Therefore,
these light curves were decomposed into the slowly varying and
rapidly varying signals with the Fourier transform. First, the power
spectrum of the light curve was calculated using the fast Fourier
transform (FFT). Second, the power spectrum was split into domains
with lower and higher frequency. Third, the inverse FFT was applied
to these two domains to obtain the slowly varying (background) and
rapidly varying (quasi-periodic peaks) signals \citep[see
also][]{Ning14a,Ning14b}. Here, a uniform timescale of 120~s was
used to distinguish these two signals for all the observation data.

\begin{figure}
\centering
\includegraphics[width=\linewidth,clip=]{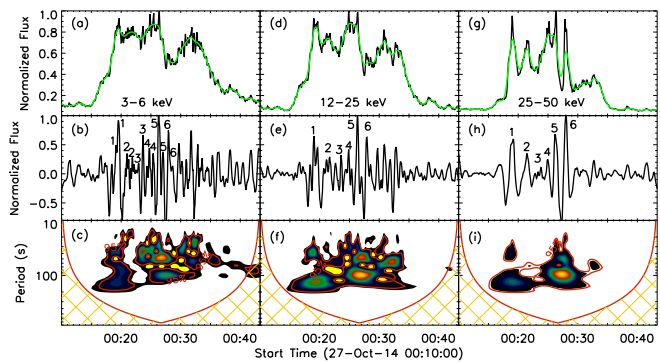}
\caption{Upper panels: {\it RHESSI} light curves (black) and their
slowly varying signals (green). Middle panels: the rapidly
varying
signals. Lower panels: the wavelet power spectra. \label{qpp1}}
\end{figure}

Figure~\ref{qpp1}~(a, d, and g) shows the slowly varying signals
(green profiles) that overlap the X-ray light curves at three energy
channels from {\it RHESSI}. The middle panels display their rapidly
varying signals. In order to make the rapidly varying signal more,
it was normalized by its slowly varying signal \cite[see
also][]{Kupriyanova10,Kupriyanova13}. The period is clearly
identified in the higher energy channel, that is, 25$-$50~keV.
Panel~(h) clearly shows 6 peaks (as marked by the numbers) from
around 00:18~UT to 00:28~UT ($\sim$600~s), which means that the
period is about 100~s at this channel. Then we applied a wavelet
analysis to the rapidly varying signals, and we show the results in
the bottom panels. The wavelet results further confirm the period of
about 100~s at the higher energy, see panel~(i). However, at the
lower energy (3$-$6~keV), there are more peaks and their intervals
are shorter, that is, 12 peaks from about 00:18~UT to 00:28~UT, and
some peaks are weaker than others. The wavelet results show a period
of $\sim$50~s, as shown in panel~(c), which is twice lower than the
period at the higher energy. It is interesting that both the periods
of $\sim$50~s and $\sim$100~s are detected in the medium-energy
channel, that is, 12$-$25~keV, as shown in panel~(f). Panel~(e) also
shows that there are 6 main peaks in the rapidly varying signals at
12$-$25~keV, as indicted by the numbers. Meanwhile, there are double
subpeaks at almost every main peak. This is possible because the
emissions from this channel include both thermal and nonthermal
components. Therefore, periods of both $\sim$50~s and $\sim$100~s
are displayed in this channel. To our knowledge, this is the first
time that changing periods in a flare are reported that change from
the thermal components (SXR emissions) to the nonthermal components
(HXR emissions) at almost the same time.

\begin{figure}
\centering
\includegraphics[width=\linewidth,clip=]{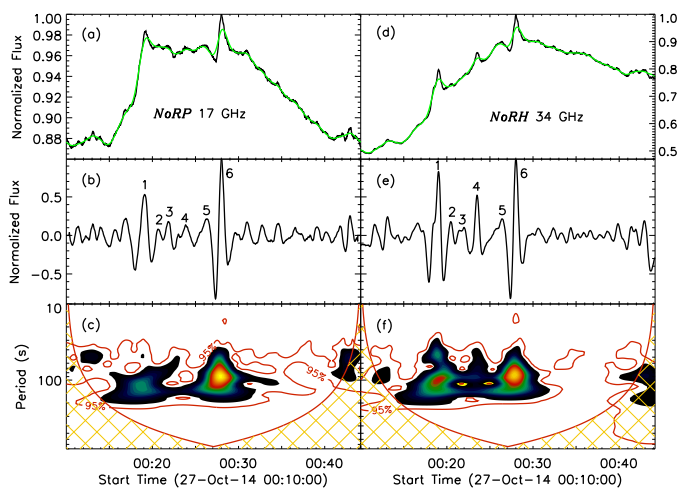}
\caption{Upper panels: Microwave emissions (black) and their
slowly varying signals (green). Middle panels: the rapidly
varying
signals. Lower panels: the wavelet power spectra. \label{qpp2}}
\end{figure}

\begin{figure}
\centering
\includegraphics[width=\linewidth,clip=]{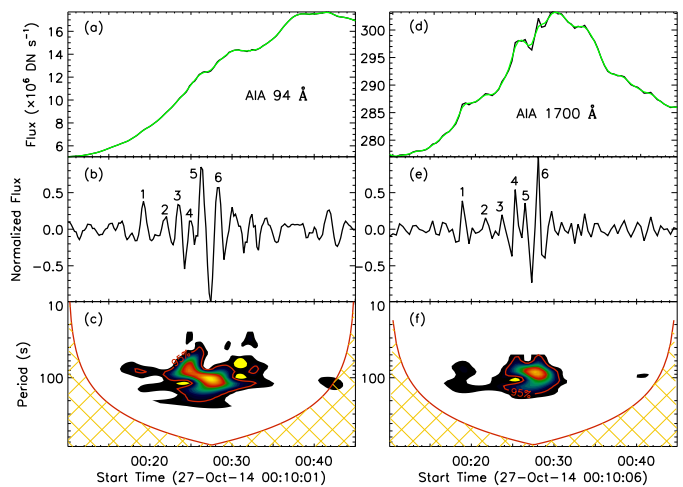}
\caption{Upper panels: EUV and UV flux (black) and their
slowly varying signals (green). Middle panels: the rapidly varying
signals. Lower panels: the wavelet power spectra. \label{qpp3}}
\end{figure}

Except for the {\it RHESSI} HXR channels, the $\sim$100~s QPPs are
also detected in microwave and EUV/UV emissions using the same
method. Figure~\ref{qpp2} shows the wavelet results from the
microwave emissions at frequencies of 17~GHz and 34~GHz. The green
profiles that overlap on the black profiles are the slowly varying
signals using the Fourier transform \citep{Ning14a,Ning14b}. The
rapidly varying signals and their wavelet power spectra are shown in
the middle and bottom panels. Similar to the {\it RHESSI} HXR
emissions, the microwave emissions at 17~GHz and 34~GHz exhibit
the $\sim$100~s QPPs. All these emissions at HXR and microwave
bands could be considered as the nonthermal components.
Figure~\ref{qpp3} shows the wavelet analysis results from the EUV/UV
wavelengths at AIA 94~{\AA} and 1700~{\AA}. In the same way as
in Fig.~\ref{qpp2}, the upper panels show the integrated intensity flux
(black profiles) and their slowly varying signals (green profiles),
and the integral region is shown in Fig.~\ref{light}~(c), while the
rapidly varying signals and their wavelet power spectra are shown in
the middle and bottom panels. The same $\sim$100~s QPPs are found
using the wavelet method. Figures~\ref{qpp2} and~\ref{qpp3} show
that there are more than 6 peaks in some channels, and some peaks
are out of the time between 00:18~UT and 00:28~UT; these are
AIA~94~{\AA} and {\it RHESSI} 12$-$25~keV. However, only 6 main
peaks are marked by the numbers from about 00:18~UT to 00:28~UT,
which is exactly the HXR peak time during the impulsive phase
of the flare.
More important, these 6 peaks appear to match well in the HXR,
microwave, and EUV/UV channels, as indicated by the number ticks.

\section{Conclusions and discussions}
We reported $\sim$100~s QPPs in the HXR, microwave, and EUV/UV
channels and the $\sim$50~s QPPs in the SXR bands in the M7.1 flare
on 2014 October 27. The result was obtained using the
multi-instrument observations from {\it RHESSI}, {\it NoRP}, {\it
NoRH}, and {\it SDO}/AIA. To our knowledge, this is the first report
about periods in the same flare that change at almost the same time
from nonthermal components to thermal components. We identified QPPs
based on the rapidly varying signals that were obtained from the
original light curves with the fast Fourier transform. Then the
$\sim$100~s and $\sim$50~s QPPs were derived with the wavelet
method. Based on this, the ratio between these two periods is 2.0.
This is well consistent with the period ratio (2.0) between the
fundamental and harmonic MHD waves.

It is very interesting that the changing periods are detected in the
same flare at almost the same time and that the periods change from
$\sim$50~s at the thermal components to $\sim$100~s at the
nonthermal components. There have been some reports about
multi-period QPPs in the same flare. However, these periods have
always been detected in the same channels, that is, in SXR light
curves \citep{Chowdhury15}, EUV flux/spectra \citep{Su12,Tian16},
H$\alpha$ images \citep{Srivastava08,Yang16}, or microwave emissions
\citep{Melnikov05,Kupriyanova14}. On the other hand, multiple
periods have been detected in various channels, but these periods
typically appeared at different times or loops, such as the
impulsive and decay phases of the flare
\citep{Zimovets10,Huang14,Hayes16,Tian16}. It is interesting that
the period ratio is exactly 2.0. This could be interpreted as
modulations in the magnetic reconnection rate that are caused by the
fundamental and harmonic modes of the MHD waves. However, previous
multi-period observations usually showed that the period ratio
deviated from 2.0 \citep[e.g.,][]{Srivastava08,Chowdhury15,Yang16},
which may be due to the longitudinal density stratification
\citep{Andries05}, the loop expansion \citep{Verth08}, the siphon
flow \citep{Li13}, or the flare phase/loop differences
\citep{Zimovets10,Tian16,Hayes16}.

It is also interesting that the QPPs exhibit the same period of
$\sim$100~s at a broad wavelength from HXR through EUV/UV to
microwave emissions. Their radiation sources spread from the
chromosphere through the transition region to the corona. It is still an
open question what modulates the 100~s QPPs at such broad wavelengths.
Three models have been presented in previous documents
\citep[e.g.,][]{Aschwanden87}: (i) MHD flux tube oscillations
modulated by waves; (ii) periodic self-organizing systems of plasma
instabilities of wave-particle or wave-wave interactions; and (iii)
modulation of periodic particle acceleration. We here observed
QPPs in
the impulsive phase, which gives us a clue about how to explain the QPPs.
It is generally thought that the magnetic reconnection process
occurs in the impulsive phase. Thus, it is possible that the QPPs
are produced by periodic magnetic reconnection in this flare. Based
on the standard flare model, magnetic reconnection could simultaneously
accelerate
the bidirectional electron beams
\citep{Aschwanden95,Su13,Li16}. The downward beam produces the
microwave emission peak in its trajectory propagation, it also
produces one peak of the HXR emission when it is injected into the
chromosphere, and this process will heat the local plasmas to
produce one peak in the EUV/UV emission. In this case, the periodic
magnetic reconnection can produce the QPPs peaks at HXR, EUV/UV, and
microwave emissions. The period of $\sim$100~s could be modulated by
certain MHD waves in the solar corona, such as the fast kink mode
\citep{Nakariakov06,Nakariakov09,Huang14}, or the slow sausage mode
\citep{Van11}, which could modulate the period of around 100~s.
However, it is difficult to determine the MHD waves in this flare
because of the observation limitation, such as the absence of
a three-dimensional magnetic configuration of the active regions.

\begin{figure}
\centering
\includegraphics[width=\linewidth,clip=]{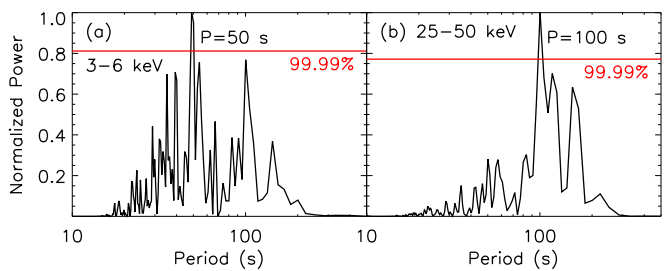}
\caption{FFT spectra of the rapidly varying signals at 3$-$6
keV, and 25$-$50 keV. The red line indicates the 99.99\% confidence
level. \label{fft1}}
\end{figure}

The wavelet spectra in Figs. 2$-$4 exhibit broad amplitudes with
brightest cores and weaker borders, which may result in a fuzzy
result for the QPP periods. To unambiguously conclude on the
periods, we performed a periodogram analysis of the rapidly varying
signals using the Lomb-Scargle periodogram method \citep{Scargle82}.
Figure~\ref{fft1} shows the FFT spectra at {\it RHESSI} 3$-$6~keV
and 25$-$50~keV, and the red line indicates the 99.99\% confidence
level defined by \cite{Horne86}. We were easily able to detect the
periods of $\sim$50~s and $\sim$100~s in the FFT spectra in the SXR
and HXR channels, respectively. This additionally confirms that the
periods change depending on the thermal or nonthermal emission.

\begin{acknowledgements}
The authors would like to thank the anonymous referee for the
valuable comments that improved the manuscript. We thank the teams of
{\it RHESSI}, {\it GOES}, {\it NoRP}, {\it NoRH}, {\it SDO}/AIA, and
{\it SDO}/HMI for their open-data use policy. This study is
supported by NSFC under grants 11603077, 11573072, 11473071,
11303101, 11333009, 973 program (2014CB744200), and Laboratory No.
2010DP173032. This work is also supported by the Youth Fund of
Jiangsu No. BK20161095 and BK20141043, and Y.~N. Su is also
supported by the one hundred talent program of the Chinese Academy of
Sciences.
\end{acknowledgements}

\end{document}